# Concept of a thermonuclear reactor based on gravity retention of high-temperature plasma

## S.I. Fisenko, I.S. Fisenko

"Rusthermosinthes" JSC
Dukat III Business Center,
6 Gasheka Str., 12th Storey,
Moscow 125047
Phone: +7 (495) 255-83-64, Fax: +7 (495) 255-83-65
E-mail: StanislavFisenko@yandex.ru

## Abstract

The fundamental fact of gravitational radiation as radiation of the same kind as the electromagnetic one has been theoretically shown in publications [1] to [6]. The obtained results are in strict correspondence to the principles of relativistic theory of gravitation and quantum mechanics. In the present paper the realization of the obtained results in relation to the dense high-temperature plasma of multivalent ions including experimental data interpretation is discussed.

## 1. Basic conception

- The problem of controlled fusion realization is directly connected to obtaining steady state of dense high-temperature plasma. It can also be unambiguously stated that the present state of the art (retaining plasma by magnetic fields of various configurations, squeezing by laser radiation) does not solve the problem of dense high-temperature plasma retention for a time required for the reaction of nuclear fusion but only solving the problem of heating plasma to the state when these reactions can exist.

In the offered method of forming dense-high temperature plasma steady states for nuclear fusion a new fundamental concept is used, namely retaining plasma by radiated gravitational field as radiation of the same kind as electromagnetic. This concept is described in details in publications [1] to [6].

The new concept of plasma retention leads to possibility of using the carbon cycle wherein there are no neutrons in the reaction products and the energy is derived from the reactor in the form of electromagnetic radiation.

## 2. A series of actions required for obtaining steady states of dense-high temperature plasma

- Forming and accelerating binary plasma with multivalent ions by accelerating magnetic field in a pulse high-current discharge.

- Injection of binary plasma from the space of the accelerating magnetic field:

exciting stationary states of an electron in its own gravitational field in the range of energy up to 171 keV with following radiation (Fig. 1) under the condition of quenching lower excited energy levels of ion electron shell of a heavy component (Fig. 2, including quenching excited state of electrons directly in nuclei of small sequential number as carbon) when retarding plasma bunch ejected from the space of the accelerating magnetic field. Cascade transitions from the upper levels are realized in the process of gravitational radiation energy transit to long-wave range.

The sequence of the operations is carried out in a two-sectional chamber of MAGO installation (Fig. 3, developed in Experimental Physics Research Institute, Sarov, [7]; the structure of the installation is most suitable for the claimed method of forming steady states of the dense high-temperature plasma) with magnetodynamic outflow of plasma and further conversion of the plasma bunch energy (in the process of quenching) in the plasma heat energy for securing both further plasma heating and exciting gravitational radiation and its transit into a long-wave part of the spectrum with consequent plasma compression in the condition of radiation blocking and increasing.

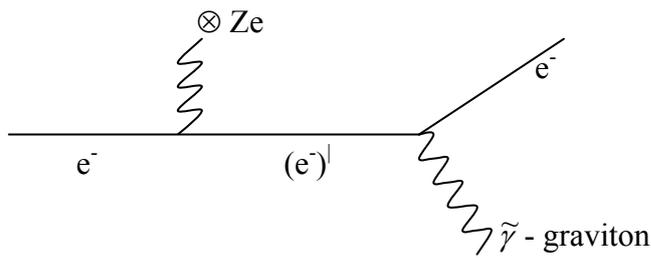

Figure 1. Graviton emission when quenching an electron in a nucleus.

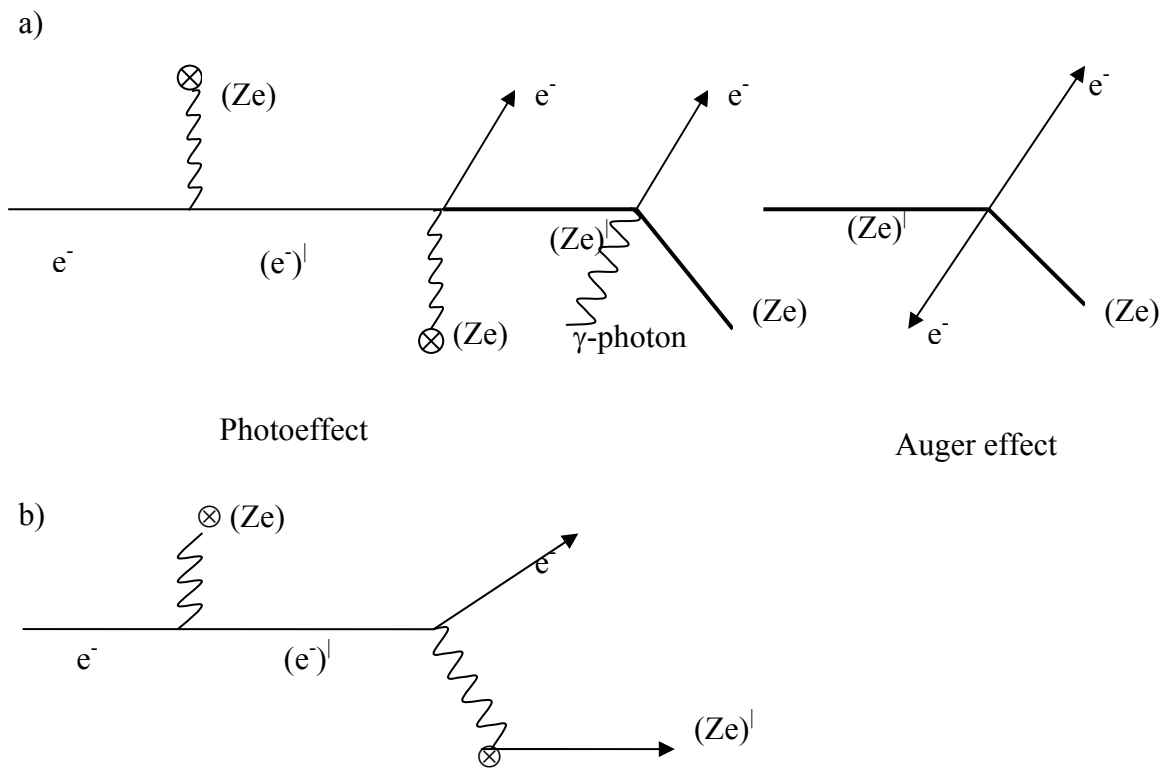

Figure 2. Quenching lower excited states of electron by: a) many-electron ions (photoelectric effect with release of one electron or autoionization (Auger effect) with release of two electrons depending on the ion number and quenching energy); b) nuclei without electron shells when an excited electron returns to normal state transferring excess energy directly to the nucleus with higher probability for the lover energy levels of excited electrons.

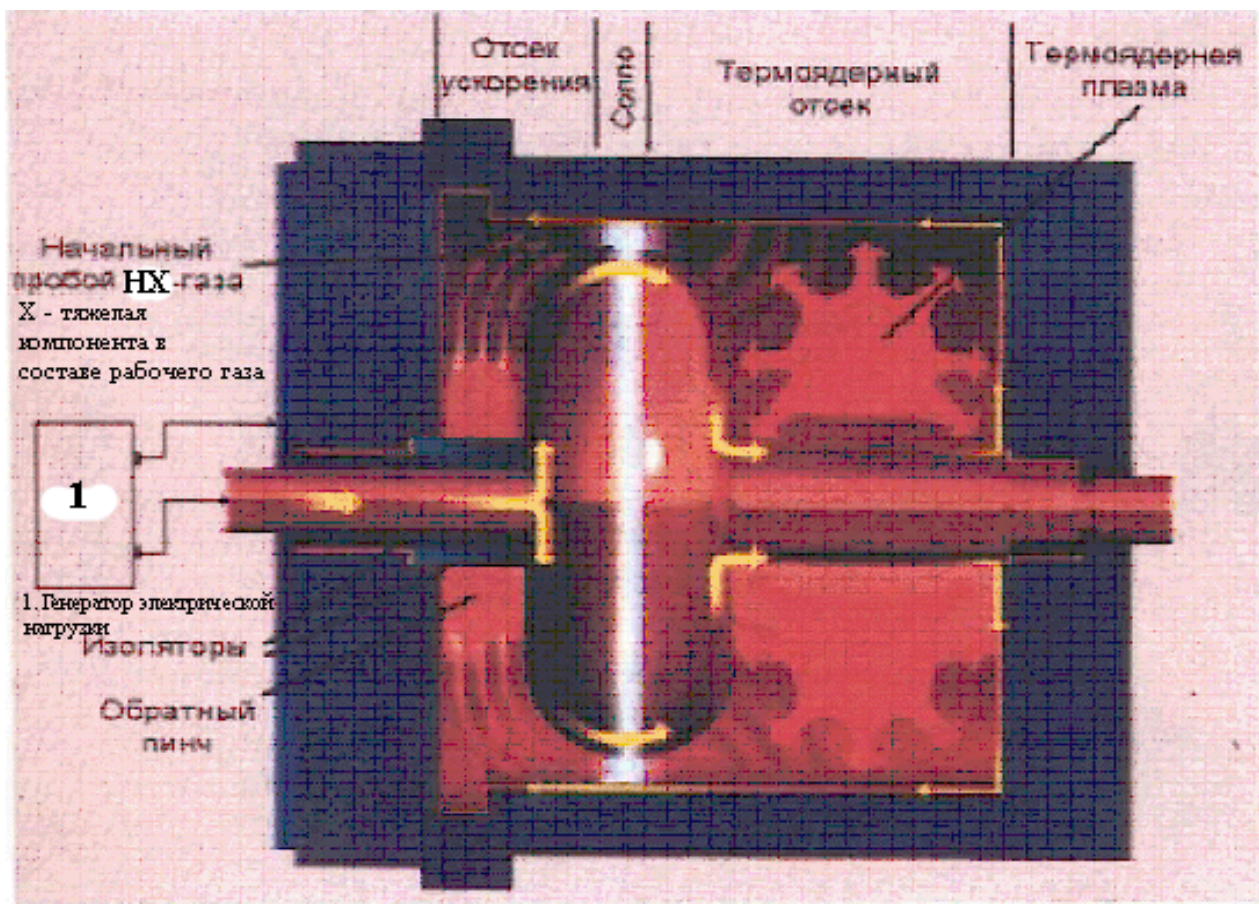

Figure 3. Physical diagram of thermonuclear plasma in MAGO chamber.

Of interest there are two modes of the installation operations depending on the work gas composition:

- a composition with hydrogen and xenon providing only for achieving steady states of plasma with consequent realization of thermonuclear reactions for compositions of (d+t) + multi-charge atoms type;

- a composition with hydrogen and carbon providing thermonuclear reactions of carbon cycle in plasma steady state mode, including energy pick-up in the form of electromagnetic radiation energy.

### 3. Experimental data

- **Registration of electron gravitational radiation lines and energy spectrum in their own gravitational field:**

It is known that the form of free neutron decay β-spectrum satisfactorily corroborates theoretical dependence for allowed transitions except soft parts of β-spectrum. Corresponding theoretical and experimental spectra are shown in Figs. 4, 5. The soft part of the spectrum is clearly linear exactly corresponding (taking into account kinetic energy of an outgoing electron) to the spectrum of electron steady states in its own gravitational field in then range of the steady state energies up to 171 keV.

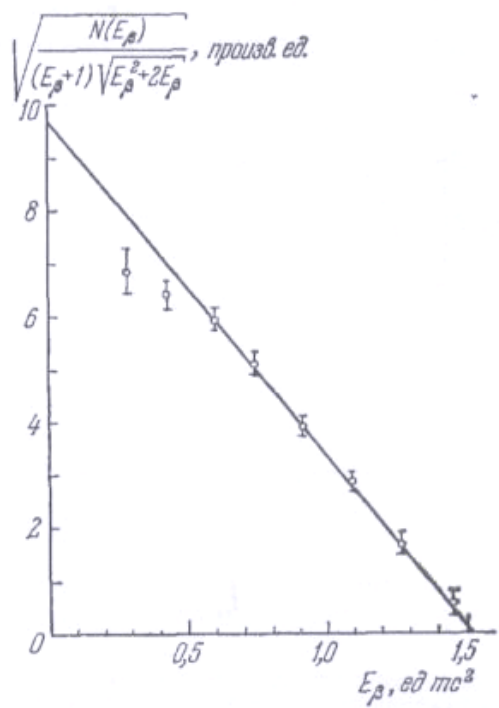
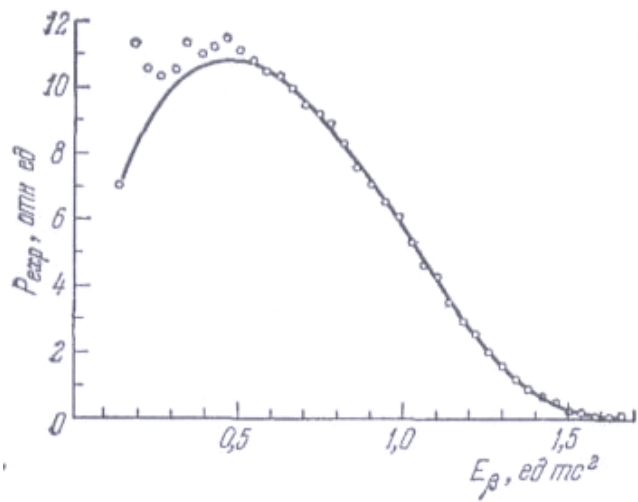

Figure 4. Beta-spectrum of free neutron decay obtained by Robson [8].

The strait line is Fermi graph, the experimental data points according to Robson [8]

Figure 5. Beta-spectrum of free neutron decay obtained by Christensen et al. [9].

The curve corresponds to a theoretical spectrum corrected for spectrometer energy resolution.

In independent experiments when at the same time electron energy distribution after electron beam passing through a foil was registered, clearly line energy spectrum was observed: Fig. 6(a). The line radiation spectrum is also clearly seen: Fig. 6(b) which can not be explained only by the presence of accelerated electron groups. The quantitative identification of the spectrum requires more precise and broad measurements including identification algorithm of energy spectrum quantitative values relating directly to steady states of electrons. Nevertheless, registered the line type of electron energy spectrum and corresponding line electron radiation spectrum preliminary corroborate as a rough approximation the very fact of electron steady states in their own gravitational field exactly in the energy range up to 171 keV.

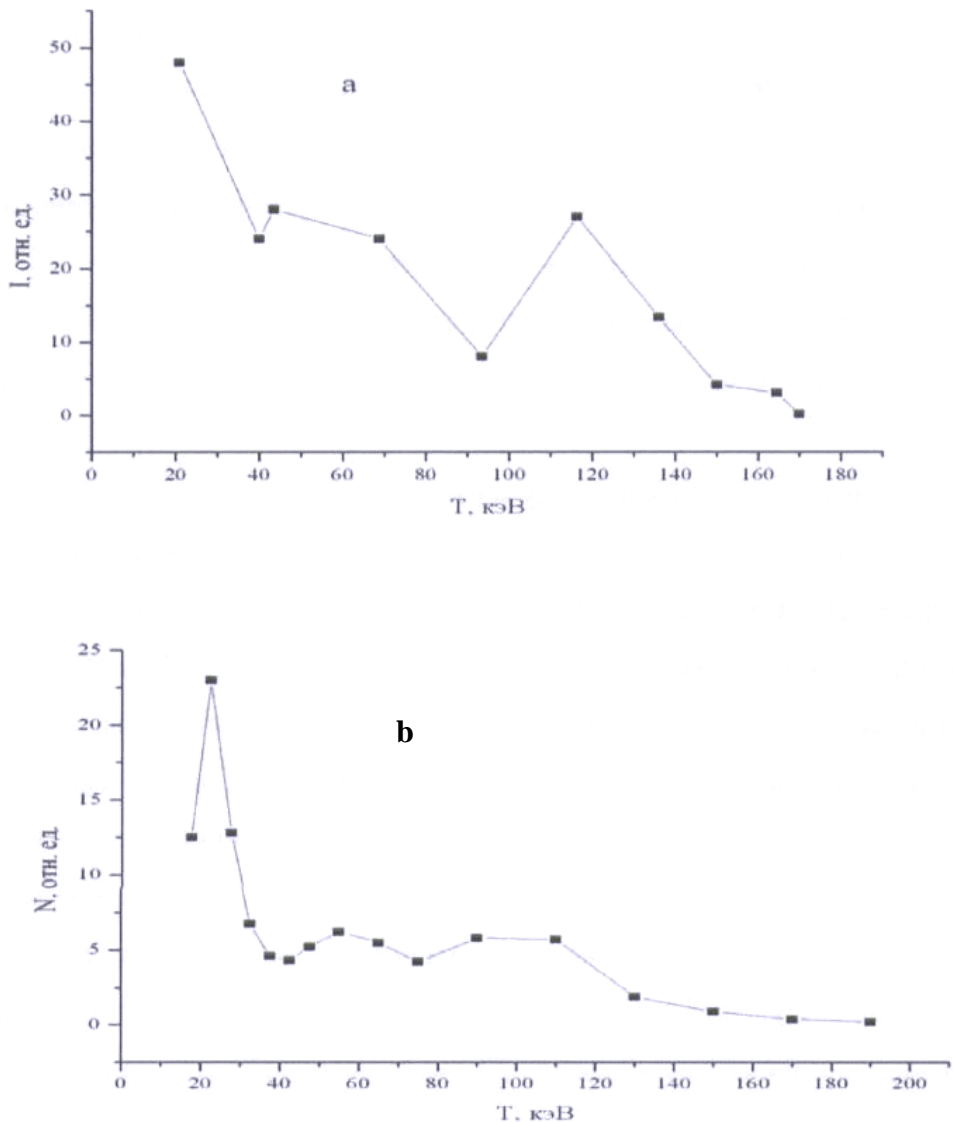

Figure 6. Energy distribution of (a) electrons and (b) X-ray quanta

In the experiments on radiation spectroscopy of channeled electrons line electromagnetic radiation spectra are registered. Typical spectra for different electron beam densities are shown in Fig. 7 wherein spectra c), d) determine the limits of the method application as a non-destructive method of crystal analysis depending on the beam density. Positions of the quasi-characteristic radiation lines are determined with the accuracy within 1% whereas their width with the accuracy within 10%. The channeled particles line spectra theory is developed rather well both quantitatively and qualitatively but conformance of design and experimental spectra is still achieved by the adjustment of averaged crystal potential.

The presence of the line spectrum of channeled electron quasi-characteristic radiation energy, including in the range of gravitational radiation energy line spectra should appear in a thinner spectrum and in widening of close spectrum lines. Factually the spectrum line widening in Fig. 7 (a, b) is an additive effect of quasi-characteristic electromagnetic and gravitational radiation of electrons. Their difference (more accurately division) is not observed at the measuring accuracy indicated above, and theoretical electromagnetic radiation spectra are simply adjusted to experimental spectra by the form of the averaged crystal potential.

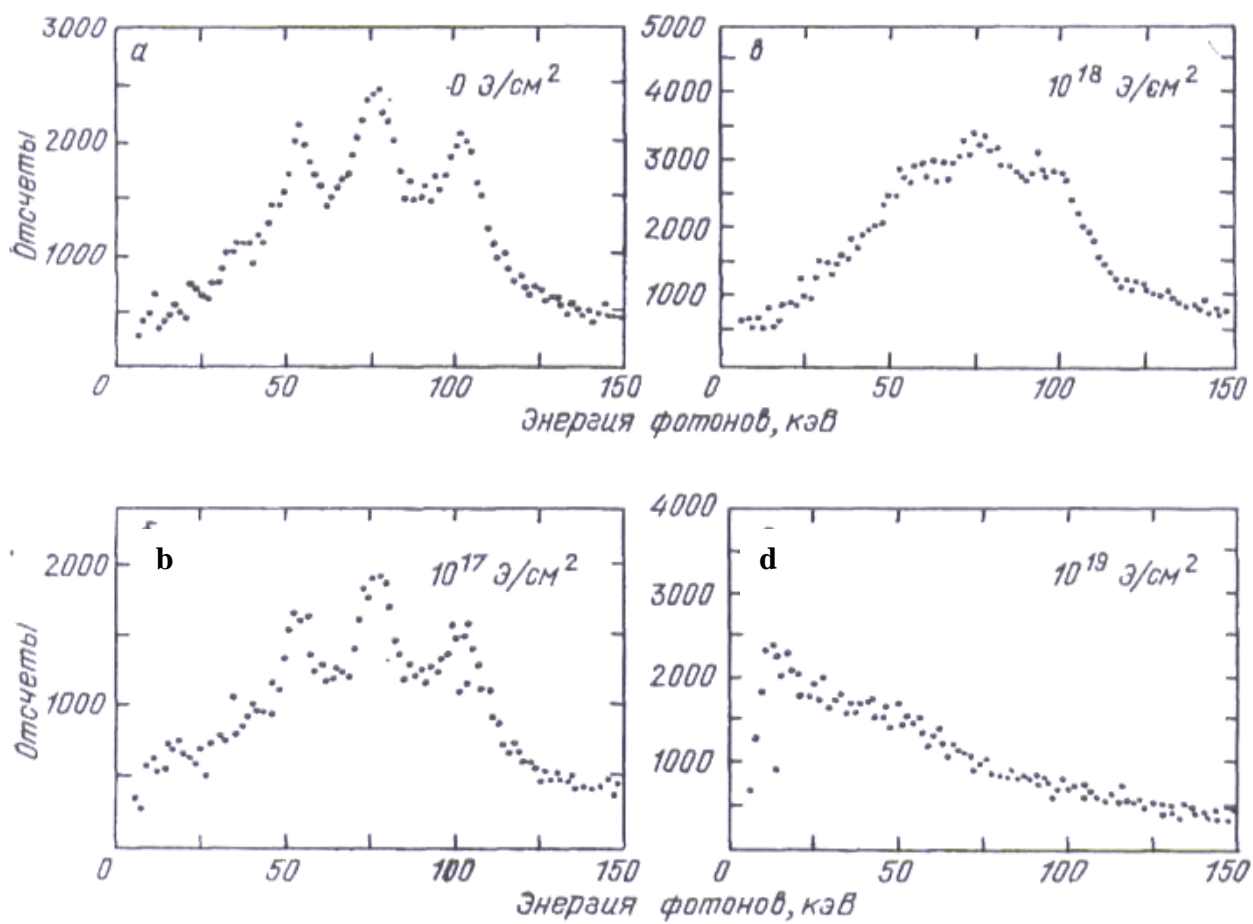

Figure 7. Radiation spectra of electrons having energy of 54 MeV in LiF monocrystall after radiation by electrons of dose 0 (a), $10^{17}$ el/cm$^2$ (b), $10^{18}$ el/cm$^2$ (c), $10^{19}$ el/cm$^2$ (d). [11].

It is obvious that these data need to be supplemented with direct experimental identification both regarding both electron gravitational radiation spectrum lines and electron steady state energy spectrum in its own gravitational field. Fig. 8 shows electron beam energy spectra in a pulse accelerator measured by a semicircular magnetic spectrometer. Two peaks of the energy spectra are connected to the feature of the pulse accelerator operations, the secondary pulse is due to lower voltage. This leads to the second (low-energy) maximum of the energy spectrum distribution.

A telemetry error in the middle and soft parts of the spectrum is not more than ± 2%. The magnetic spectrometer was used for measuring the energy spectrum of electrons after passing through the accelerator anode grid and also spectra of electrons after passing though a foil arranged behind the accelerator mesh anode. These data (and the calculated spectrum) are presented in Fig. 8. Similar measurements were carried out for Ti foil (foil thickness 50 μm) and Ta (foil thickness 10 μm). In case of Ti the measurements were limited from the top by energy of 0.148 MeV, and in case of TA by energy of 0.168 MeV. Above these values the measurement errors increase substantially (for this type of the accelerator). The difference between the normalized spectral densities of theoretical and experimental electron spectra after passing through Ti, Ta and Al foils, Fig. 9. The data indicate that there is a spectrum of electron energy states in their own gravitational field when the electrons are excited when passing through a foil. The obtained data are not sufficient for numerical spectrum identification but the very fact of the spectrum presence according to the data is doubtless.

ΔN$_e$/ΔE, electron/MeV

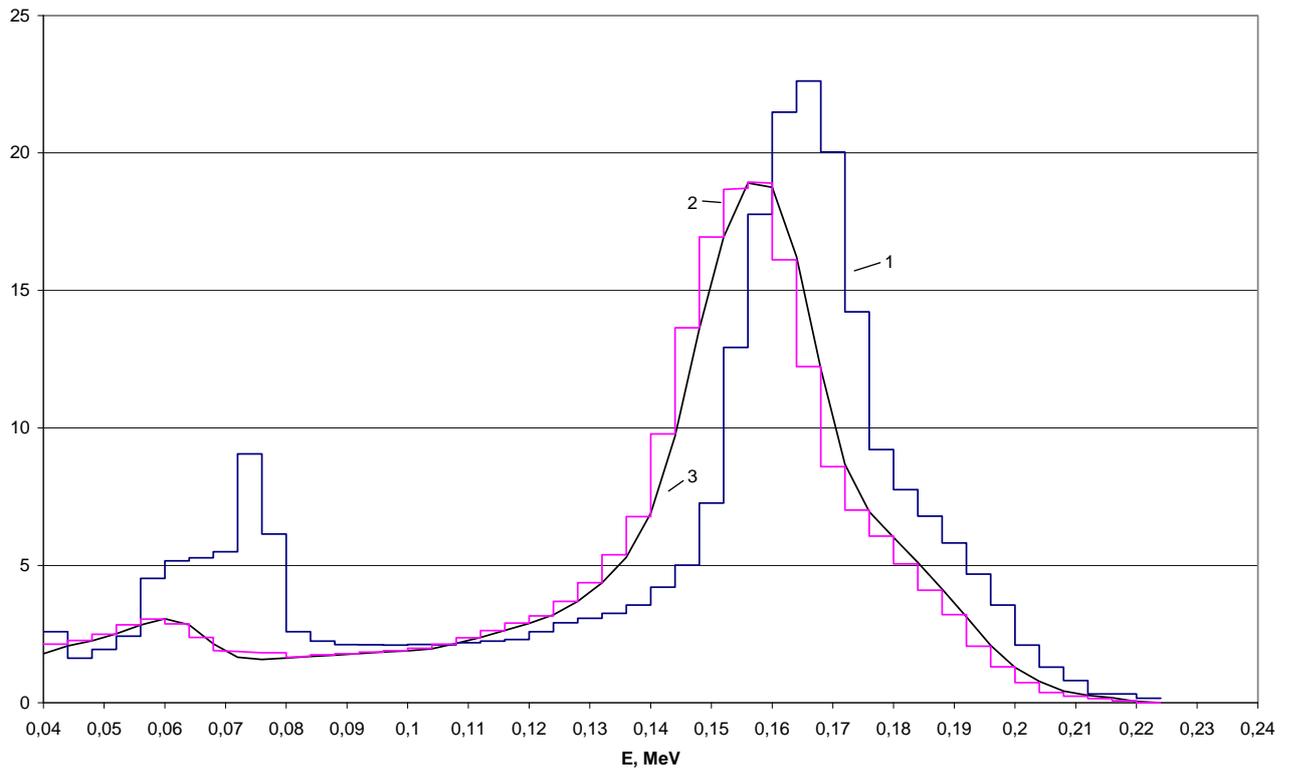

Figure 8. Electron energy spectra: 1 – after passing the grid, 2 – after passing the Al foil 13 μm thick; 3 – spectrum calculation according to ELIZA program based on the database [12] for each spectrum 1. The spectrum is normalized by the standard.

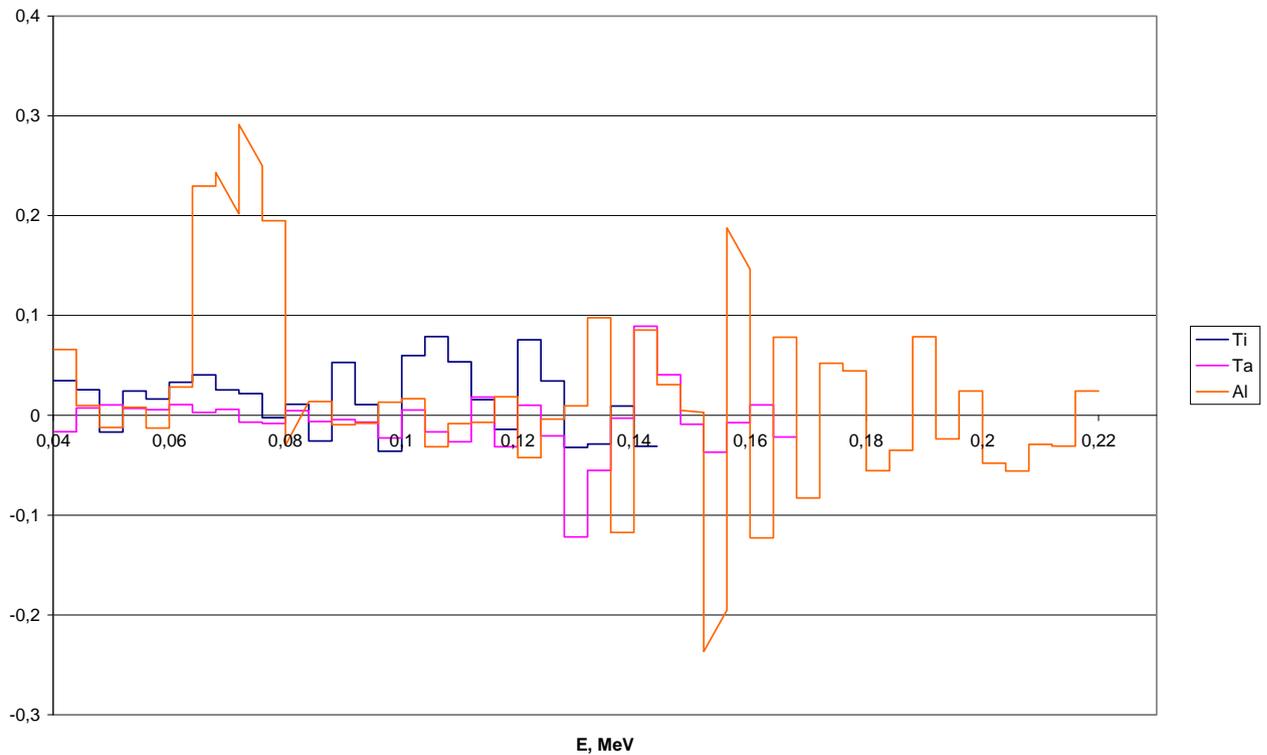

Figure 9. Difference of spectral density for theoretical and experimental spectra of electrons passed through Ti, Ta and Al foils.

- **Micropinch plasma electron gravitational radiation in pulse high-current discharges**

The concept of a thermonuclear reactor on the principle of compressing dense high-temperature plasma by emitted gravitational field is supported by the processes of micropinching multicharged ion plasma in pulse high-current diodes. Figs 10, 11(a) show characteristic parts of micropinch soft X-ray radiation spectrum. Micropinch spectrum line widening does not correspond to existing electromagnetic conceptions but corresponds to such plasma thermodynamic states which can only be obtained with the help of compression by gravitational field, radiation flashes of which takes place during plasma thermalization in a discharge local space. Such statement is based on the comparison of experimental and expected parts of the spectrum shown in Fig. 11. Adjustment of the expected spectrum portion to the experimental one [14] was made by selecting average values of density $\rho$, electron temperature $T_e$ and velocity gradient $\nabla U$ of the substance hydrodynamic motion.

As a mechanism of spectrum lines widening, a Doppler, radiation and impact widening were considered. Such adjustment according to said widening mechanisms does not lead to complete reproduction of the registered part of the micropinch radiation spectrum. This is the evidence (under the condition of independent conformation of the macroscopic parameters adjustment) of additional widening mechanism existence due to electron excited states and corresponding gravitational radiation spectrum part already not having clearly expressed lines because of energy transfer in the spectrum to the long-wave area.

That is to say that the additional mechanism of spectral lines widening of the characteristic electromagnetic radiation of multiple-charge ions (in the conditions of plasma compression by radiated gravitational field) is the only and unequivocal way of quenching electrons excited states at the radiating energy levels of ions and exciting these levels by gravitational radiation at resonance frequencies. Such increase in probability of ion transitions in other states results in additional spectral lines widening of the characteristic radiation. The reason for quick degradation of micropinches in various pulse high-currency discharges with      multiple-charge ions is also clear. There is only partial thermolization of accelerated plasma with the power of gravitational radiation not sufficient for maintaining steady states.

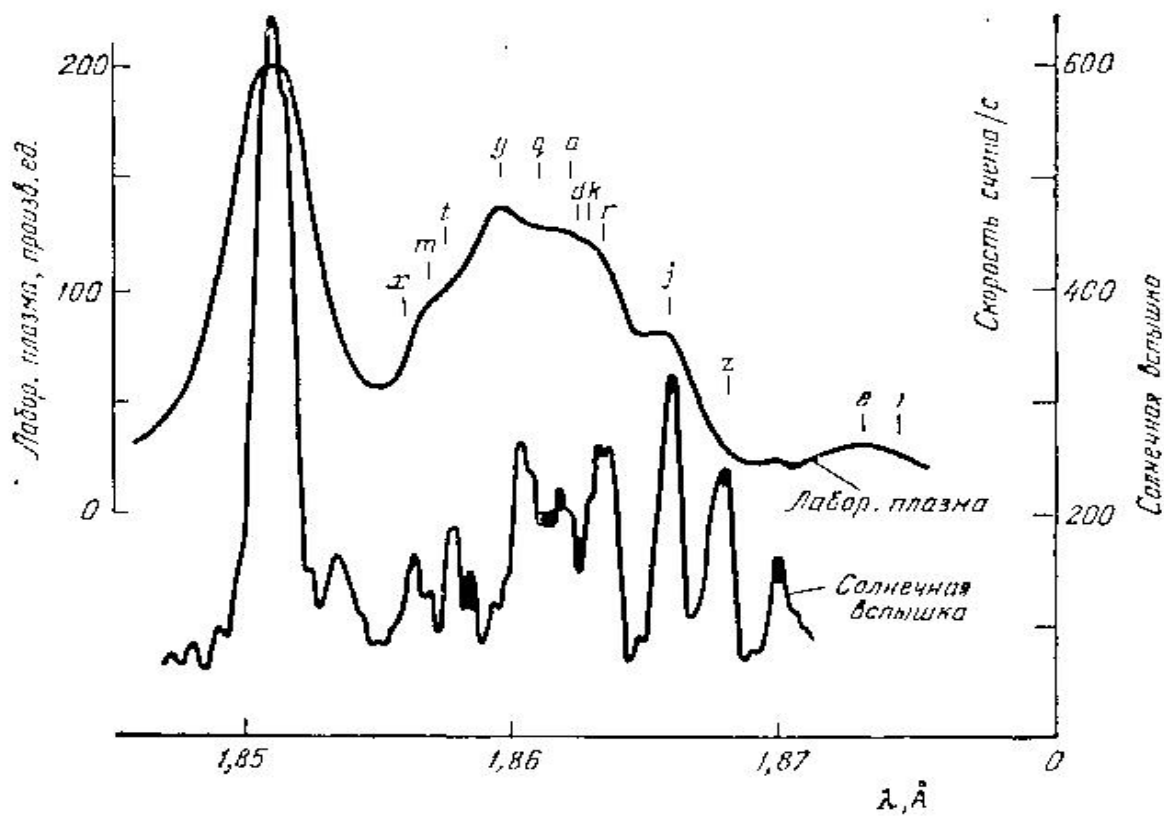

Figure 10.  A part of vacuum sparkle spectrum and a corresponding part of solar flare spectrum. [13].

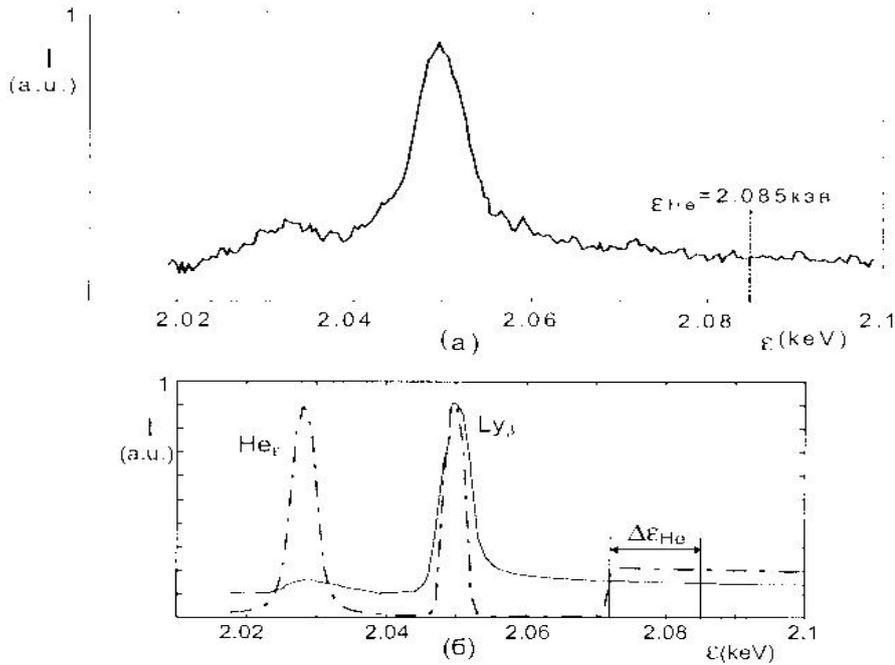

Figure 11. Experimental (a) and calculated (b) parts of a micropinch spectrum normalized for line Lyβ intensity in the area of the basic state ionization threshold of He-like ions.
The firm line in variant (b) corresponds to density of 0.1 g/cm$^3$, the dotted line – to 0.01 g/cm$^3$; it was assumed that $T_e$ = 0.35 keV, [14].

### 3. Thermonuclear plasma steady states generation:

Available experimental data show that they can be reproduced in an active experiment directly in the MAGO chamber on the ground of existence of gravitational radiation narrow-band spectrum in the range up to 171 keV with long-wave spectrum part realized by cascade electron transitions from the upper energy levels. Quenching lower excited states of electrons on the electron shell energy levels of heavy component ions in combination with cascade transitions will result in plasma compression in the conditions of blockage and gravitational radiation intensification.

The MAGO chamber capacity of work with deuterium-tritium composition was tested experimentally. The obtained experimental data of plasma compression in MAGO chamber (including data obtained in the framework together with the Los-Alamos National Laboratory [15]) prove that there is the fusion reaction (Fig. 12); however the holding time is not sufficient, there need to be longer holding time. The choice of such design as a design for a thermonuclear reactor is unequivocal since it is completely corresponds to the system of exciting and amplifying gravitational radiation when plasma is thermolized after outflow from the nozzle, and required additional compression actually takes place when the working plasma composition is changed (many-electron ions) and current-voltage characteristic of the charge changes correspondingly. The simplicity of the MAGO chamber technical structure is even more clearly shown by the possibility to use as the generator of electrical load such devices as a capacitors battery or an autonomous magnetic explosion generator (VMG) with all consequences of practical use of such thermonuclear reactor.

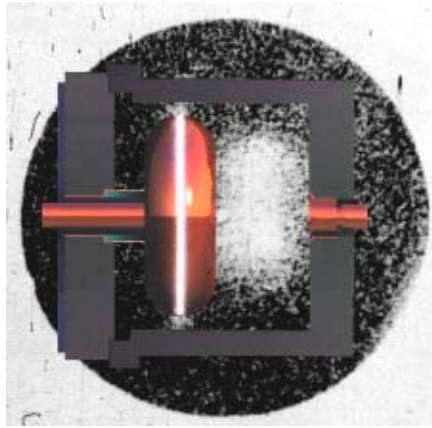

Figure 12. Neutron picture of the neutron generation area.

**Conclusion**

1. A concept of a thermonuclear reactor based on the method of forming plasma steady states based on the idea of holding by the radiated gravitational field as the radiation of the same level as the gravitational one is disclosed.

2. Feasibility of the concept is supported by the data which are incomplete but not conflicting with the uniform system.